\documentclass[aps,twocolumn,superscriptaddress,showpacs]{revtex4}
\usepackage{graphicx}
\usepackage{dcolumn}
\usepackage{amssymb}
\usepackage{bm}
\newcommand\mib[1]{\bm{#1}}

\def\({\left (}
\def\){\right)}
\def\[{\left [}
\def\]{\right]}
\def\<{\left <}
\def\>{\right>}

\def\+{\!+\!}
\def\-{\!-\!}

\begin{document}

\title{
    On-Line Learning Theory of Soft Committee Machines\\
    with Correlated Hidden Units\\
    -- Steepest Gradient Descent and Natural Gradient Descent --
}

\author{Masato Inoue}
\affiliation{
    Laboratory for Mathematical Neuroscience, Brain Science Institute, RIKEN\\
    2-1, Hirosawa, Wako, Saitama 351-0198, Japan\\
    E-mail: \{minoue, hypark, okada\}@brain.riken.go.jp
}
\affiliation{
    Department of Otolaryngology - Head and Neck Surgery, Graduate School of Medicine, Kyoto University\\
    54, Kawara-cho, Shogoin, Sakyo-ku, Kyoto 606-8507, Japan
}

\author{Hyeyoung Park}
\affiliation{
    Laboratory for Mathematical Neuroscience, Brain Science Institute, RIKEN\\
    2-1, Hirosawa, Wako, Saitama 351-0198, Japan\\
    E-mail: \{minoue, hypark, okada\}@brain.riken.go.jp
}

\author{Masato Okada}
\affiliation{
    Laboratory for Mathematical Neuroscience, Brain Science Institute, RIKEN\\
    2-1, Hirosawa, Wako, Saitama 351-0198, Japan\\
    E-mail: \{minoue, hypark, okada\}@brain.riken.go.jp
}

\begin{abstract}
     The permutation symmetry of the hidden units in multilayer perceptrons causes the saddle structure and plateaus of the learning dynamics in gradient learning methods. The correlation of the weight vectors of hidden units in a teacher network is thought to affect this saddle structure, resulting in a prolonged learning time, but this mechanism is still unclear. In this paper, we discuss it with regard to soft committee machines and on-line learning using statistical mechanics. Conventional gradient descent needs more time to break the symmetry as the correlation of the teacher weight vectors rises. On the other hand, no plateaus occur with natural gradient descent regardless of the correlation for the limit of a low learning rate. Analytical results support these dynamics around the saddle point.
\end{abstract}


\maketitle

\section{Introduction}
     One of the biggest problems of neural network learning is the plateau of the learning curve. Considering the gradient learning method and its generalization error, this plateau is mainly caused by the saddle structure of the error function. The permutation symmetry prevents the identification of the hidden units in multilayer perceptrons if they have the same weight vectors, and produces this saddle structure \cite{rf:1,rf:2}. In the learning scenario of a teacher and a student network, the saddle is thought to be affected by the strength of the correlation of the hidden units in the teacher network, which may be closely related to the length of the plateau. More specifically, in the conventional gradient descent (GD), the weight vectors in the student network are known to approach the saddle before reaching their final states \cite{rf:2}. Since the saddle is located between the weight vectors of the teacher hidden units, their stronger correlation is supposed to force the student weight vectors closer to the saddle, resulting in a longer plateau.

     Natural gradient descent (NGD), however, may be able to avoid the saddle because it can update the network parameters to the optimal direction in the Riemannian space \cite{rf:3}. NGD is a fairly general method for effectively adjusting the parameters of stochastic models, but its validity in multilayer perceptrons is uncertain because of three intrinsic problems: 1) NGD needs prior knowledge of the input distribution to calculate the Fisher information matrix, 2) NGD is unstable around the singular points of the Fisher information matrix, 3) matrix inversion is time consuming, which might be critical especially in real-time learning. The method proposed by Yang and Amari\cite{rf:4} can be used to calculate NGD efficiently in the case of a large input dimension in multilayer perceptrons. Also, the adaptive method can be used to approximate the inverse of the Fisher information matrix asymptotically without prior knowledge or matrix inversion \cite{rf:5}. In this paper, we discuss the problem of singularity; since the saddle is one of the singular points, how NGD works around there is one of our main topics. 

     On-line learning is one of the most popular forms of training. Analysis of the network dynamics in on-line learning is much easier than for batch learning because the state of the network and the learning samples are independent of each other. In this framework, the statistical mechanics method proposed by Saad and Solla can be used to analyze the GD dynamics exactly at the large limit of the input dimension \cite{rf:2}. Rattray and Saad extended this technique to NGD and reported that it works efficiently in multilayer perceptrons \cite{rf:6}. In this paper, we also use this method and contrast the dynamics for GD and NGD, focusing on the corrupted saddle structure under a strong correlation of the hidden units in the teacher network.

\section{Model}
\begin{figure}[t]
\begin{center}
    \includegraphics[width=85mm]{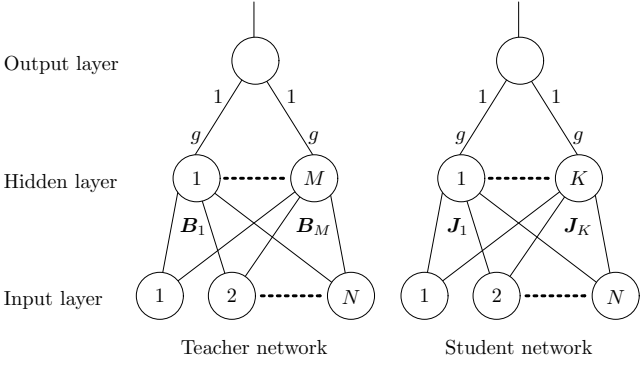}
    \caption{Teacher and student networks. Each weight between any hidden unit and the output is fixed to 1.}
    \label{fig:1}
\end{center} 
\end{figure}
     Soft committee machines (Fig. \ref{fig:1}) are considered where the teacher network has $M$ hidden units while the student has $K$ units. To apply NGD, Gaussian noise $n \sim {\mathcal N}(0, \sigma^2)$ is added to the output of the student;
\begin{eqnarray}
    \label{eq:1}
    \zeta &\equiv& f\mib{_B}(\xi), \;\;\;\;\;\;\;\;
    f\mib{_B}(\mib\xi) \equiv \sum_{k=1}^{M}{g(\mib{B}_k^T \mib\xi)},\\
    \label{eq:2}
    \zeta' &\equiv& f\mib{_J}(\xi)+n,\quad
    f\mib{_J}(\mib\xi) \equiv \sum_{k=1}^{K}{g(\mib{J}_k^T \mib\xi)},
\end{eqnarray}
where $\mib\xi\in{\mathbb R}^N$ denotes the input vector while $\mib B_i\in{\mathbb R}^N$ and $\mib J_i\in{\mathbb R}^N$ are the $i$th weight vectors of the teacher and the student networks, respectively. Here, $T$ means the transposition while $g$ is an activation function.

     The joint probability distribution of the input $\mib \xi$ and the output $\zeta'$ of the student network is given by
\begin{eqnarray}
    \label{eq:3}
    & p\mib{_J}(\mib\xi,\zeta') \equiv p(\mib\xi)p\mib{_J}(\zeta'|\mib\xi), &\\
    \label{eq:4}
    & p\mib{_J}(\zeta'|\mib\xi) \equiv \frac{1}{\sqrt{2\pi\sigma^2}}
    \exp{ \( -\frac{\{\zeta'-f\mib{_J}(\mib\xi)\}^2}{2\sigma^2} \) }. &
\end{eqnarray}
The parameter vector of (\ref{eq:3}), $\mib{J} \equiv [\mib J_1^T, \mib J_2^T,..., \mib J_K^T ]^T \in {\mathbb R}^{KN}$ , is updated iteratively to approximate the joint probability distribution of the input $\mib\xi$ and the output $\zeta$ of the teacher network,
\begin{eqnarray}
    \label{eq:5}
    p(\mib\xi,\zeta) \equiv p(\mib\xi)\delta(\zeta-f\mib{_B}(\mib\xi)),
\end{eqnarray}
where $\delta$ is the delta function. The loss function for a given set of a learning sample $\{\mib\xi,\zeta\}$, defined using the logarithmic loss of the conditional probability distribution of (\ref{eq:4}), is
\begin{eqnarray}
    \label{eq:6}
    \epsilon\mib{_J}(\mib\xi,\zeta) \equiv -\ln p\mib{_J}(\zeta|\mib\xi)+c_0 = \frac{1}{2\sigma^2}\{\zeta-f\mib{_J}(\xi)\}^2,
\end{eqnarray}
where $c_0 \equiv -\ln{\sqrt{2\pi\sigma^2}}$ is constant. The generalization error is then defined as the expected loss:
\begin{eqnarray}
    \label{eq:7}
    \epsilon_g(\mib J) \equiv \< \epsilon\mib{_J}(\mib\xi,\zeta) \> \!_\{ \!\mib{_\xi} \,\!_{,\zeta\}}.
\end{eqnarray}
The definitions of (\ref{eq:6}) can be written, by applying (\ref{eq:1}) and (\ref{eq:2}), as
\begin{eqnarray}
    \label{eq:8}
    \epsilon\mib{_J}(\mib\xi,\zeta) = \epsilon\mib{_J}(\mib\xi) \equiv \frac{1}{2\sigma^2}
    \left \{ f\mib{_B}(\mib\xi) - f\mib{_J}(\mib\xi) \right\}^2.
\end{eqnarray}

     We consider on-line learning in this paper, where the parameter vector $\mib J$ is updated for each set of an independently given sample $\{\mib\xi,\zeta\}$. The updating rule, the differential of $\mib J$, for GD is defined with a learning rate $\eta$ as
\begin{eqnarray}
    \label{eq:9}
    \Delta\mib J = -\frac{\eta}{N}\nabla \mib{_J} \epsilon\mib{_J} (\mib\xi,\zeta),
\end{eqnarray}
where 
\begin{eqnarray}
    \label{eq:10}
    \nabla \mib{_J}\,\!_i \epsilon\mib{_J} (\mib\xi,\zeta) = -\frac{1}{\sigma^2}g'(\mib{J}_i^T \mib\xi)
    \left \{ f\mib{_B}(\mib\xi) - f\mib{_J}(\mib\xi) \right\} \mib\xi,
\end{eqnarray}
where $g'$ denotes the derivative of $g$. One for NGD is also defined as
\begin{eqnarray}
    \label{eq:11}
    \Delta\mib J = -\frac{\eta}{N} \mib G^{-1} \nabla \mib{_J} \epsilon\mib{_J} (\mib\xi,\zeta),
\end{eqnarray}
where $\mib G$ denotes the Fisher information matrix of the parameter vector $\mib J$:
\begin{eqnarray}
    \label{eq:12}
    \mib G \equiv \< [\nabla \mib{_J} \ln{p \mib{_J} (\mib\xi,\zeta')}] [\nabla \mib{_J} \ln{p \mib{_J} (\mib\xi,\zeta')}]^T \> \!_\{\!\mib{_\xi} \,\!_{,\zeta'\}}.
\end{eqnarray}
The $\mib G$ can be written, in block form, as
\begin{eqnarray}
    \label{eq:13}
    & \mib G = \left[\!\! \begin{array}{ccc}
        \mib G_{1,1} & \cdots & \mib G_{1,K} \\
        \vdots & \ddots & \vdots \\
        \mib G_{K,1} & \cdots & \mib G_{K,K} \\
    \end{array} \!\!\right]\!\! ,&\nonumber\\
    & \mib G_{ij} = \frac{1}{\sigma^2} \< g'(\mib{J}_i^T \mib\xi) g'(\mib{J}_j^T \mib\xi) \mib\xi \mib\xi^T \> \,\!\!_\{\!\mib{_\xi} \!_{\}}.&
\end{eqnarray}
In the case of the standard multivariate normal distribution input, $\mib\xi \sim {\mathcal N}(\mib 0, \mib I)$, the inverse of the Fisher information matrix is also given by
\begin{eqnarray}
    \label{eq:14}
    & \mib G^{-1} = \left[\!\! \begin{array}{ccc}
        \mib G_{1,1}^{-1} & \cdots & \mib G_{1,K}^{-1} \\
        \vdots & \ddots & \vdots \\
        \mib G_{K,1}^{-1} & \cdots & \mib G_{K,K}^{-1} \\
    \end{array} \!\!\right]\!\!, &\nonumber\\
    & \mib G_{ij}^{-1} = \sigma^2 \{ \theta_{ij} \mib I + \mib J' \mib \Theta_{ij} \mib J'\,\!^T \}, &
\end{eqnarray}
where $\mib J' \equiv [ \mib J_1,..., \mib J_K ]$ is a $N$ by $K$ matrix, while $\theta_{ij}$ is a scalar and $\mib \Theta_{ij}$ is a $K$ by $K$ matrix \cite{rf:4}.

\section{Theory}
\subsection{Order parameters and generalization error}
     At the thermodynamics limit, the limit of $N\to\infty$, the dynamics of the network can be analyzed using statistical mechanics. Here, the order parameters that represent the correlations of the weight vectors are used instead of the $N$-dimensional vectors $\mib\xi$, $\mib B_i$, and $\mib J_i$. To make the present paper self-contained, we briefly summarize the derivation of the order parameter equations of the soft committee machine \cite{rf:2,rf:6}.

     From here on, the input vector is assumed to obey a $N$-dimensional multivariate Gaussian noise with zero mean and a unit covariance matrix: $\mib\xi \sim {\mathcal N}(\mib 0, \mib I)$. The correlation between the input and each weight vector, denoted by $x_i \equiv \mib{J}_i^T \mib\xi$ and $y_i \equiv \mib{B}_i^T \mib\xi$, is then distributed as a normal distribution; $x_i \sim {\mathcal N}(0, \mib{J}_i^T \mib{J}_i)$ and $y_i \sim {\mathcal N}(0, \mib{B}_i^T \mib{B}_i)$, while each covariance of them is given by $\<x_i x_j\> \!_\{\!\mib{_\xi}\!_{\}} = \mib{J}_i^T \mib{J}_j$, $\<x_i y_j\> \!_\{\!\mib{_\xi}\!_{\}} = \mib{J}_i^T \mib{B}_j$, and $\<y_i y_j\> \!_\{\!\mib{_\xi}\!_{\}} = \mib{B}_i^T \mib{B}_j$. Therefore, a new vector, defined as
\begin{eqnarray}
    \label{eq:15}
    \mib z \equiv [x_1,...,x_K,y_1,...,y_M]^T \in {\mathbb R}^{K+M},
\end{eqnarray}
is distributed as a multivariate normal distribution ${\mathcal N}(\mib 0, \mib C)$:
\begin{eqnarray}
    \label{eq:16}
    p(\mib z) = \frac{1}{\sqrt{\{2\pi\}^{K+M}|\mib C|}} \exp{\( -\frac{1}{2} \mib z^T \mib C^{-1} \mib z \)},
\end{eqnarray}
where $\mib C$ is the variance-covariance matrix:
\begin{eqnarray}
    \label{eq:17}
    \mib C \equiv \left[\!\! \begin{array}{cc} \mib Q & \mib R \\ \mib R^T & \mib T \end{array} \!\!\right]\!\!
\end{eqnarray}
with
\begin{eqnarray}
    \label{eq:18}
    \mib Q \equiv \mib J'\,\!^T \mib J' &=& \left[\!\! \begin{array}{ccc}
        Q_{1,1} & \cdots & Q_{1,K} \\
        \vdots & \ddots & \vdots \\
        Q_{K,1} & \cdots & Q_{K,K} \\
    \end{array} \!\!\right]\!\!,\\
    \label{eq:19}
    \mib R \equiv \mib J'\,\!^T \mib B' &=& \left[\!\! \begin{array}{ccc}
        R_{1,1} & \cdots & R_{1,M} \\
        \vdots & \ddots & \vdots \\
        R_{K,1} & \cdots & R_{K,M} \\
    \end{array} \!\!\right]\!\!,\\
    \label{eq:20}
    \mib T \equiv \mib B'\,\!^T \mib B' &=& \left[\!\! \begin{array}{ccc}
        T_{1,1} & \cdots & T_{1,M} \\
        \vdots & \ddots & \vdots \\
        T_{M,1} & \cdots & T_{M,M} \\
    \end{array} \!\!\right]\!\!,
\end{eqnarray}
and $\mib J' \equiv [ \mib J_1 \cdots \mib J_K ]$, $\mib B' \equiv [ \mib B_1 \cdots \mib B_M ]$. Here, $\mib Q$ and $\mib R$ are the order parameters of this system.

     Using these order parameters, the generalization error in (\ref{eq:7}), $\epsilon_g(\mib J) \equiv \< \epsilon\mib{_J}(\mib\xi,\zeta) \> \!_\{ \!\mib{_\xi} \,\!_{,\zeta\}}$, can be calculated by
\begin{eqnarray}
    \label{eq:21}
    \epsilon_g(\mib J) &=& \int{d\mib z\; p(\mib z) \frac{1}{2\sigma^2} \left \{ \sum_{k=1}^{M}{g(y_k)} - \sum_{k=1}^{K}{g(x_k)} \right\}^2}.
\end{eqnarray}
If we define the activation function $g$ as $g(x)=\mathrm{erf}(x/\sqrt2)$ from here on, the generalization error is given by
\begin{eqnarray}
    \label{eq:22}
    \epsilon_g(\mib J) &=& \frac{1}{\pi\sigma^2} \left \{
        -2      \sum_{i=1}^{K}{ \sum_{j=1}^{M}{ \arcsin{\frac{R_{ij}}{\sqrt{Q_{ii}\+1}\sqrt{T_{jj}\+1}}} }} \right.\nonumber\\
         &+&    \sum_{i,j=1}^{K}              { \arcsin{\frac{Q_{ij}}{\sqrt{Q_{ii}\+1}\sqrt{Q_{jj}\+1}}} } \nonumber\\
         &+&\left. \sum_{i,j=1}^{M}              { \arcsin{\frac{T_{ij}}{\sqrt{T_{ii}\+1}\sqrt{T_{jj}\+1}}} }
    \right\},
\end{eqnarray}
which depends on only the order parameters. 

\subsection{Dynamics of the order parameters}
     Here we substitute the dynamics of the order parameters for those of the system. First, we can replace the updating rule (\ref{eq:9}) with
\begin{eqnarray}
    \label{eq:23}
    \Delta\mib J_i = -\frac{\eta}{\sigma^2 N}\delta_i \mib\xi,
\end{eqnarray}
where
\begin{eqnarray}
    \label{eq:24}
    \delta_i \equiv -g'(x_i) \left \{ \sum_{k=1}^{M}{g(y_k)} - \sum_{k=1}^{K}{g(x_k)} \right\}.
\end{eqnarray}
Thus, the updating rule of the order parameters is given by
\begin{eqnarray}
    \label{eq:25}
    \Delta R_{ij} &=& [\mib J_i + \Delta\mib J_i]^T \mib B_j - \mib J_i^T \mib B_j \nonumber\\
              &=& -\frac{\eta}{\sigma^2 N}\delta_i y_j,
\end{eqnarray}
and
\begin{eqnarray}
    \label{eq:26}
    \Delta Q_{ij} &=& [\mib J_i + \Delta\mib J_i]^T [\mib J_j + \Delta\mib J_j] - \mib J_i^T\mib J_j \nonumber\\
               &=& -\frac{\eta}{\sigma^2 N} \{\delta_i x_j + \delta_j x_i\} + \frac{\eta^2}{\sigma^4 N^2}\delta_i \delta_j \mib\xi^T \mib\xi.
\end{eqnarray}

     Here we introduce the time $\alpha$; a short period, $\Delta\alpha = 1/N$, is defined to be consumed for each learning iteration. At the large limit of $N$, the differential dynamics of $R_{ij}$ and $Q_{ij}$ are calculated as 
\begin{eqnarray}
    \label{eq:27}
    \frac{\partial R_{ij}}{\partial\alpha} &=& \lim_{\Delta\alpha\to0}{\frac{\Delta R_{ij}}{\Delta\alpha}}
                               =  \lim_{N\to\infty}{N\Delta R_{ij}}
                               =  -\frac{\eta}{\sigma^2} \< \delta_i y_j \> \!_\{\!\mib{_z}\!_{\}} \nonumber\\
                              &=& -\frac{\eta}{\sigma^2} \psi_{ij}
\end{eqnarray}
and
\begin{eqnarray}
    \label{eq:28}
    \frac{\partial Q_{ij}}{\partial\alpha} = -\frac{\eta}{\sigma^2} \{\phi_{ij} + \phi_{ji}\} + \frac{\eta^2}{\sigma^4} \upsilon_{ij},
\end{eqnarray}
where $\mib\xi^T \mib\xi \to N$ is applied, while the new variables are defined as
\begin{eqnarray}
    \label{eq:29}
    \psi_{ij} \equiv \< \delta_i   y_j \> \!_\{\!\mib{_z}\!_{\}},\quad
    \phi_{ij} \equiv \< \delta_i   x_j \> \!_\{\!\mib{_z}\!_{\}},\quad
    \upsilon_{ij} \equiv \< \delta_i \delta_j \> \!_\{\!\mib{_z}\!_{\}},\quad
\end{eqnarray}
The dynamics for NGD can be provided in the same way: 
\begin{eqnarray}
    \label{eq:30}
    \Delta\mib J_i = -\frac{\eta}{\sigma^2 N} \sum_{k=1}^{K}{\delta_k \mib G_{ik}^{-1} \mib\xi}.
\end{eqnarray}
Thus, the dynamics of the order parameters are
\begin{eqnarray}
    \label{eq:31}
    \frac{\partial R_{ij}}{\partial\alpha} = -\eta \sum_{k=1}^{K}{\{ \theta_{ik} \psi_{kj} + \phi_{k \bullet} \mib\Theta_{ki} R_{\bullet j} \}},
\end{eqnarray}
\begin{eqnarray}
    \label{eq:32}
    \frac{\partial Q_{ij}}{\partial\alpha}
        &=& -\eta \sum_{k=1}^{K}{\{ \theta_{ik} \phi_{kj} \+ \theta_{jk}\phi_{ki} \+ \phi_{k \bullet} \mib\Theta_{ik}^T Q_{\bullet j} } \nonumber\\
        & & + \phi_{k \bullet} \mib\Theta_{jk}^T Q_{\bullet i} \} + \eta^2 \sum_{k,l=1}^{K}{ \theta_{ik} \theta_{jl} \upsilon_{kl}} ,
\end{eqnarray}
where $\phi_{k \bullet}$ denotes the $k$th row of the matrix $\{\phi_{ij}\} _{i,j=1,...,K}$, while $R_{\bullet j}$ denotes the $j$th column of the matrix $\mib R$, and so on \cite{rf:6}.

\section{Numerical Results}
\begin{figure*}[t]
    \includegraphics[width=58mm]{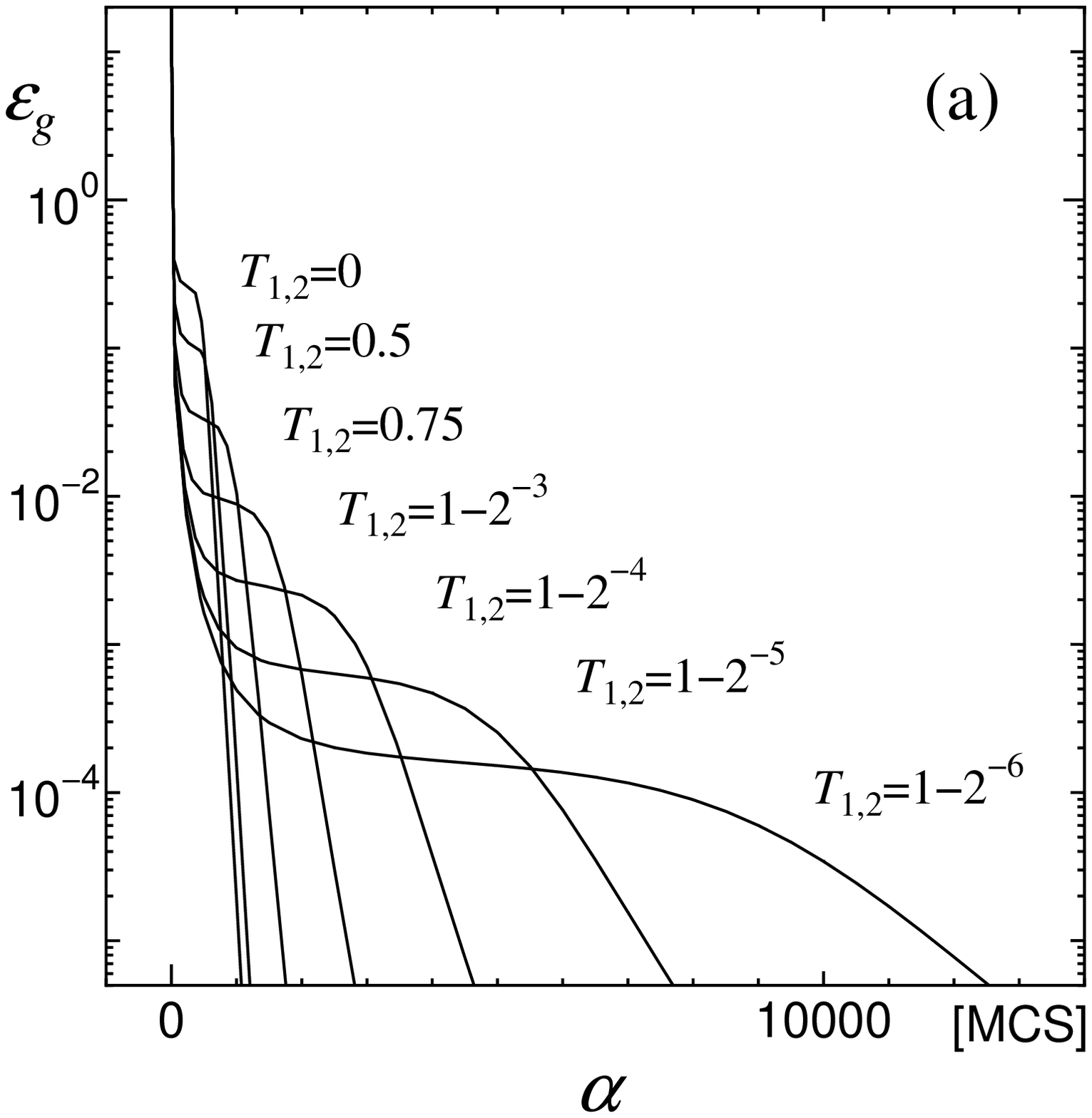}
    \includegraphics[width=58mm]{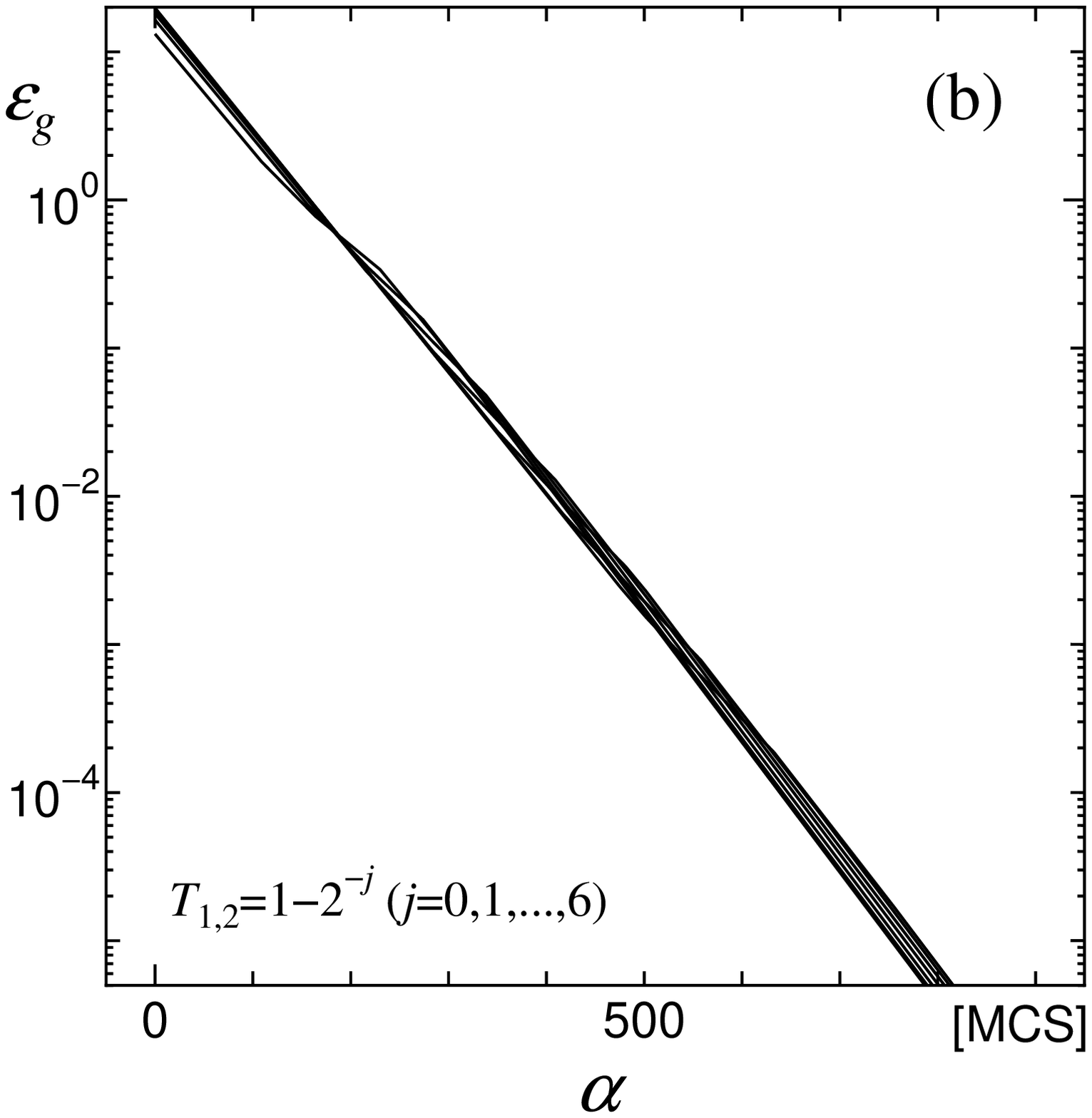}
    \includegraphics[width=58mm]{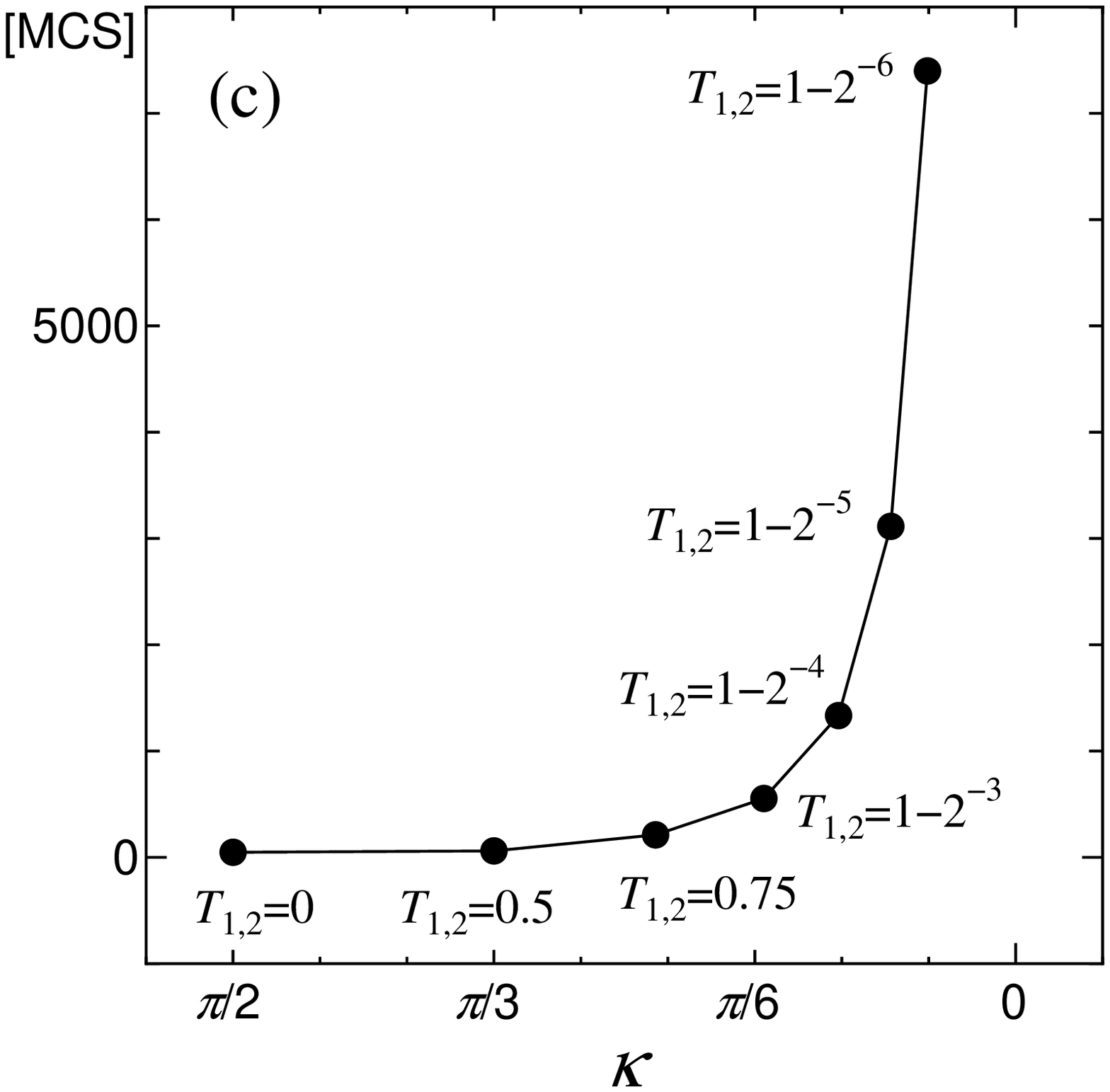}
    \caption{Time evolution of the generalization error in GD (a) and NGD (b). All the trajectories are almost completely overlapped in (b). The plateau periods in (a) were measured and are shown in (c). MCS denotes the number of Monte Carlo steps.}
    \label{fig:2}
    \vspace{24pt}
    \includegraphics[width=58mm]{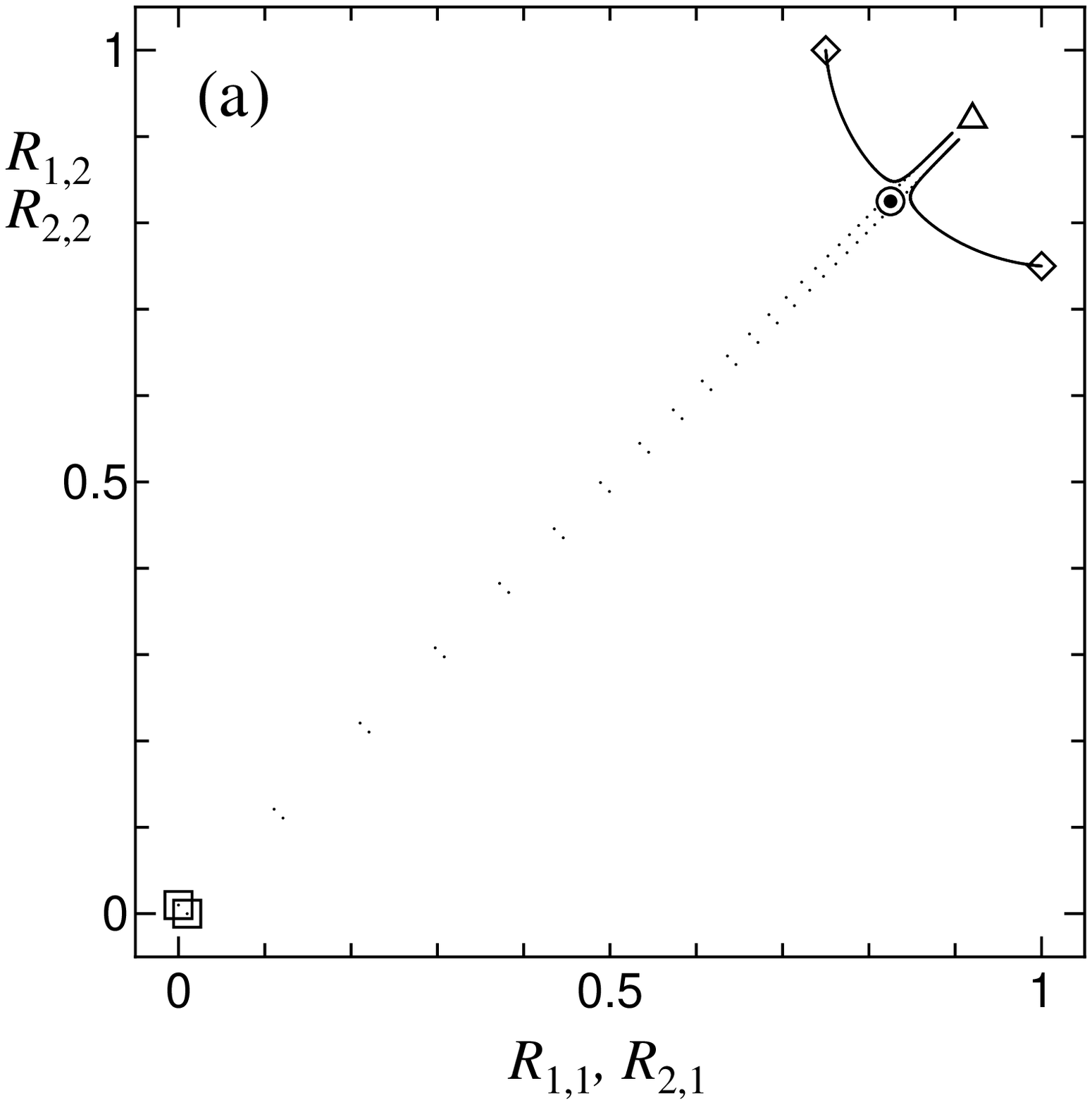}
    \includegraphics[width=58mm]{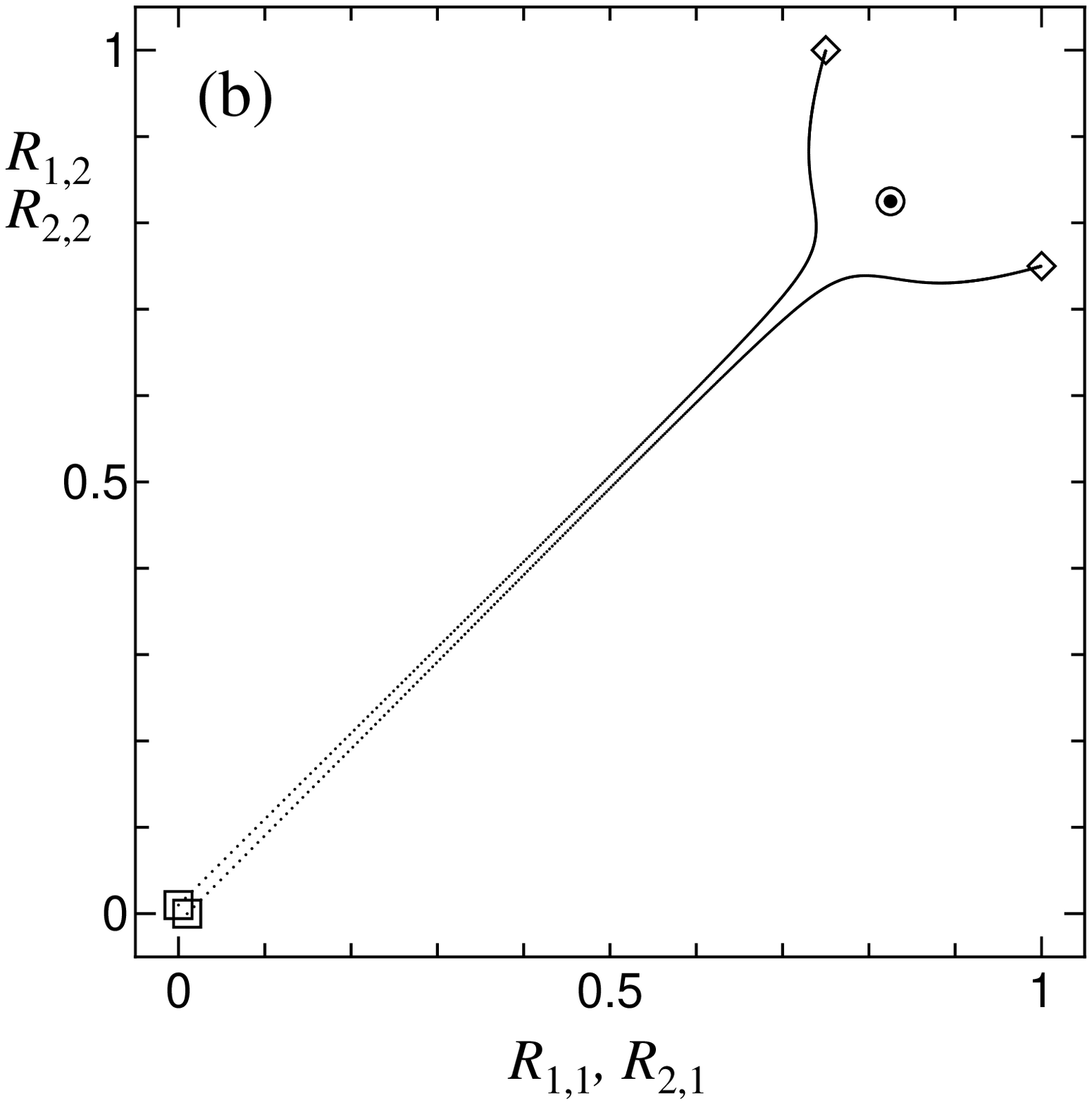}
    \caption{Time evolution of the order parameters $(R_{1,1}, R_{1,2})$ and $(R_{2,1}, R_{2,2})$ in GD (a) and NGD (b). The correlation of the teacher weight vectors $T_{1,2} = 0.75$. Start points: $\square$; turning points: $\triangle$; the saddle: $\odot$; and goals: $\diamondsuit$ .}
    \label{fig:3}
\end{figure*}

\begin{figure*}[t]
    \includegraphics[width=58mm]{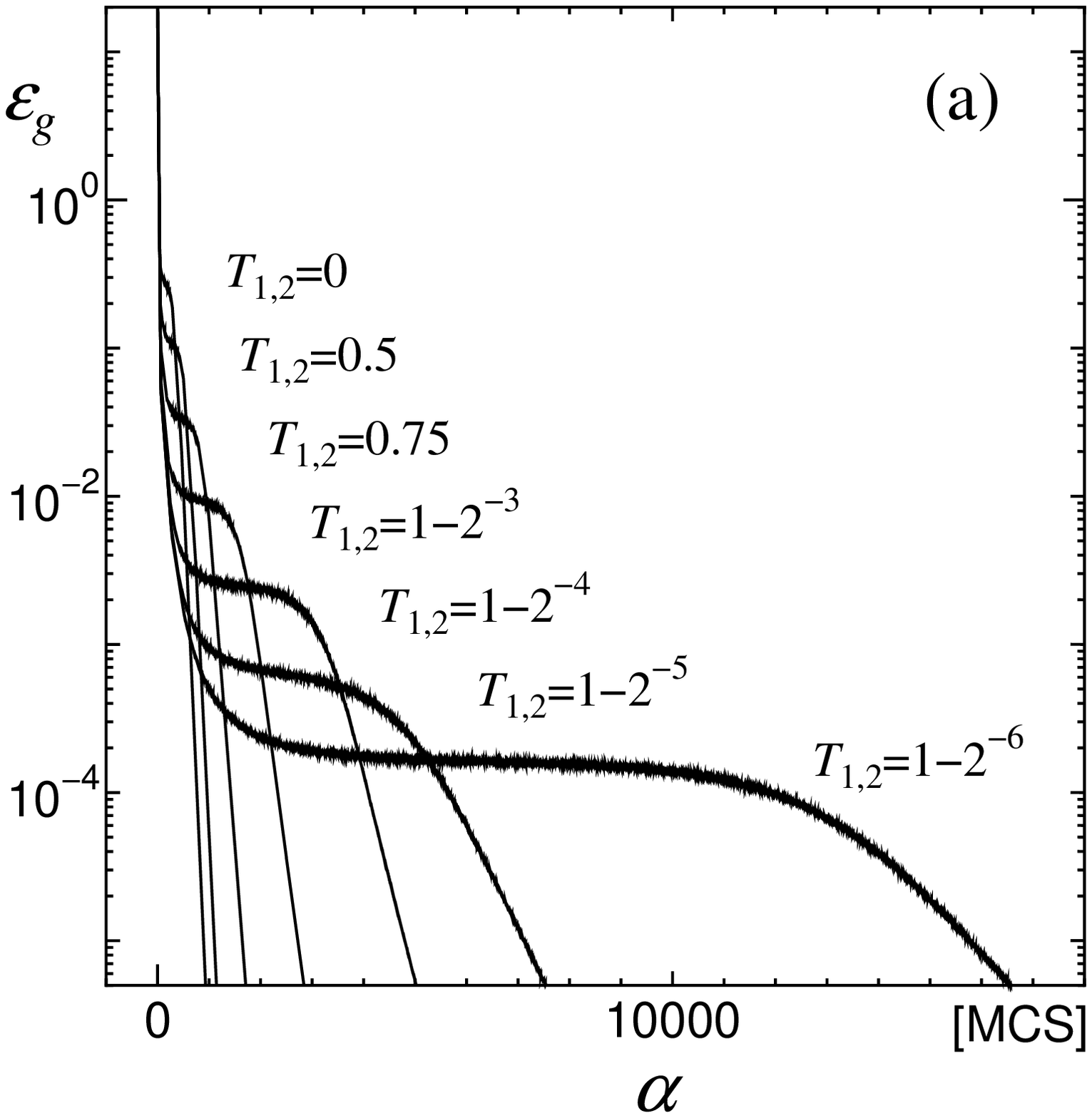}
    \includegraphics[width=58mm]{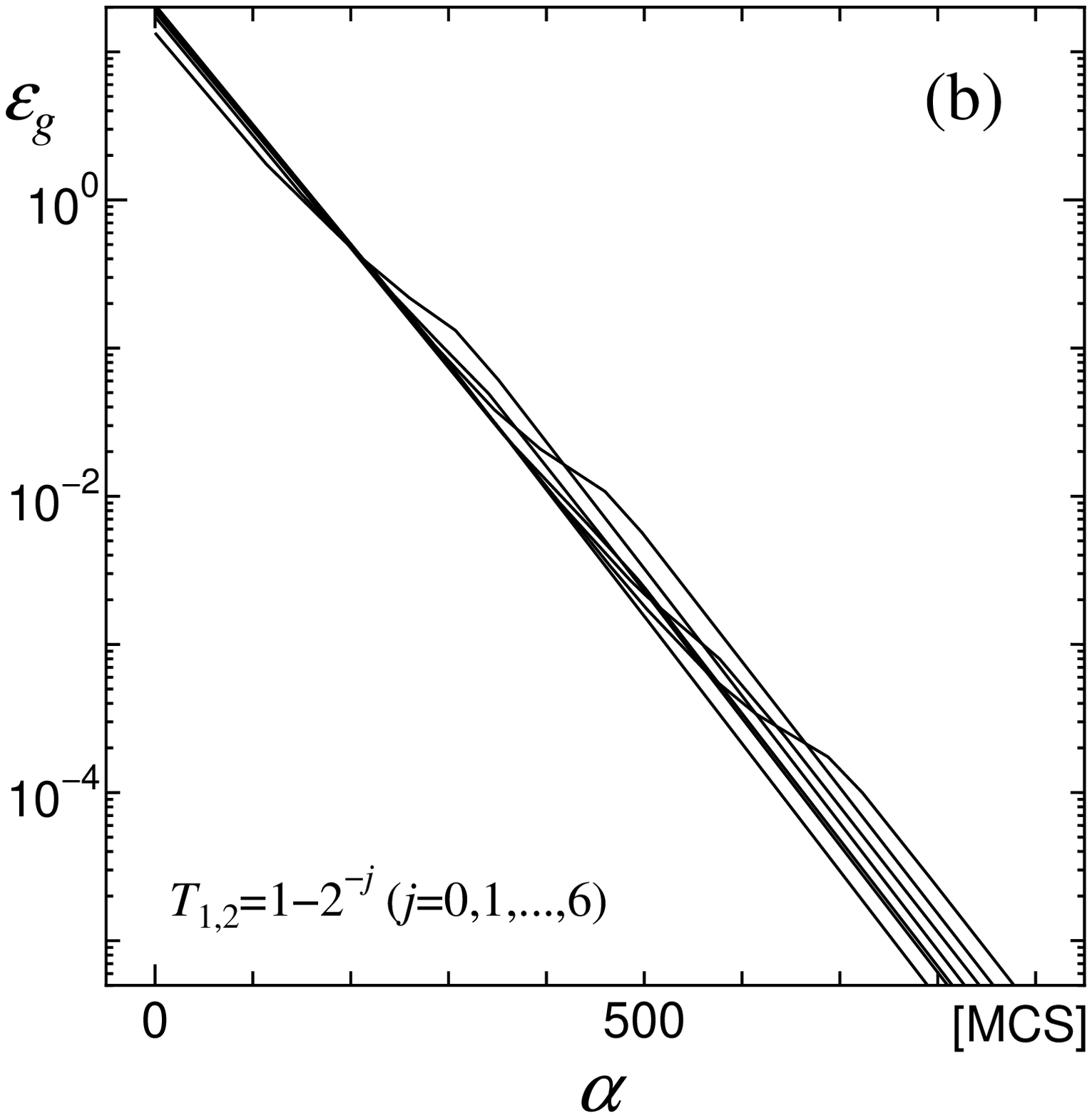}
    \caption{Time evolution of the generalization error for the case of $N=1000$ in GD (a) and NGD (b).}
    \label{fig:4}
    \vspace{24pt}
    \includegraphics[width=58mm]{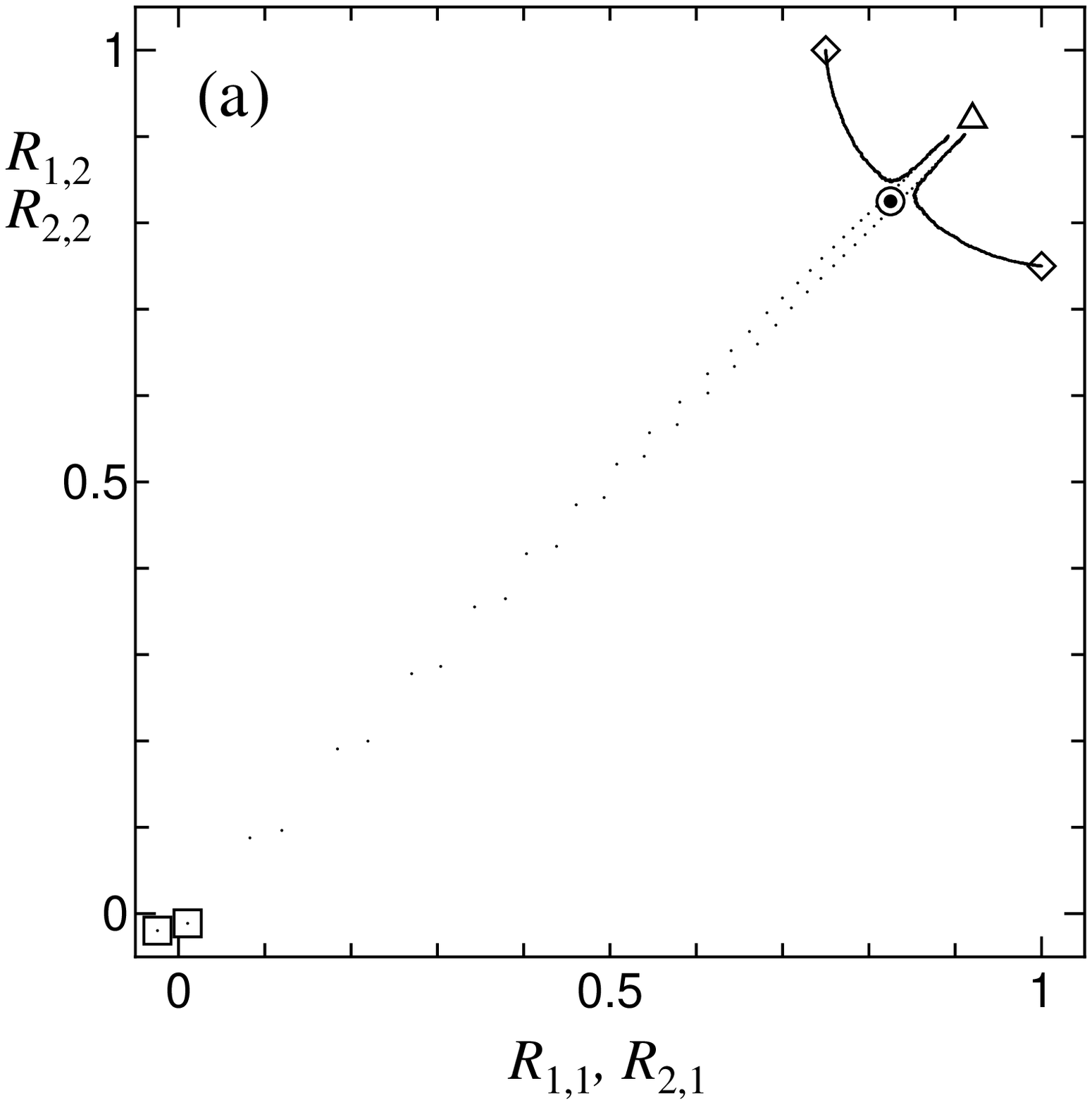}
    \includegraphics[width=58mm]{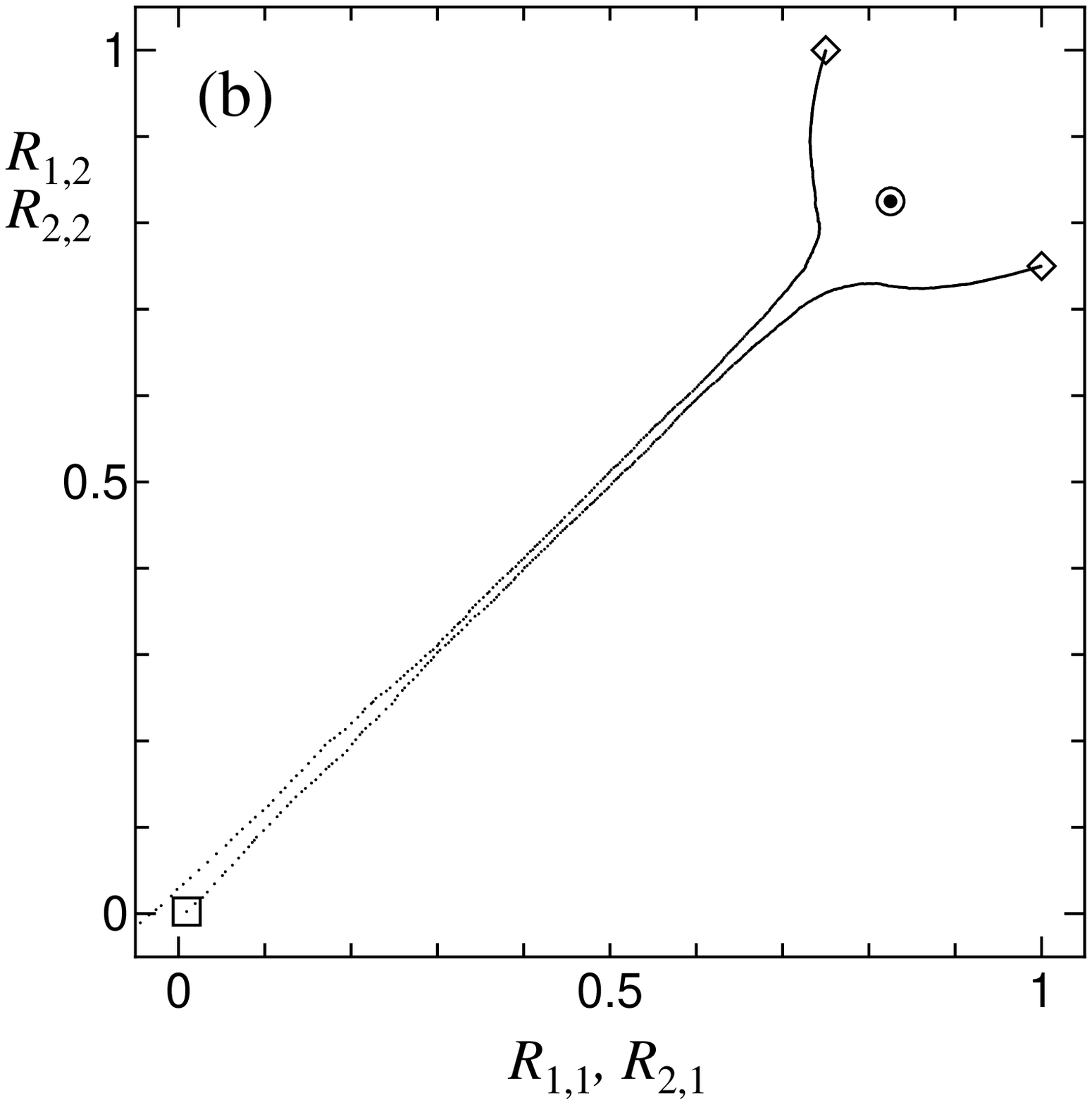}
    \caption{Time evolution of the order parameters $(R_{1,1}, R_{1,2})$ and $(R_{2,1}, R_{2,2})$ for the case of $N=1000$ in GD (a) and NGD (b) as in Fig. \ref{fig:3}.}
    \label{fig:5}
\end{figure*}
     In this section, we discuss how the learning dynamics depend on the correlation of teacher weight vectors $T_{1,2}$. The results are also contrasted between GD and NGD.

     We set the number of the hidden units and the lengths of the teacher weight vectors as follows:
\begin{eqnarray}
    \label{eq:33}
    K=M=2,\quad T_{1,1}=T_{2,2}.
\end{eqnarray}
We also restrict the initial conditions to
\begin{eqnarray}
    \label{eq:34}
    Q_{1,1}=Q_{2,2},\quad R_{1,1}=R_{2,2},\quad R_{1,2}=R_{2,1}.
\end{eqnarray}
Because of the symmetry of the system, these restrictions are preserved throughout the learning. Specifically, we use
\begin{eqnarray}
    \label{eq:35}
    \mib Q = \left[\!\! \begin{array}{cc} 1 & 0 \\ 0 & 1 \end{array} \!\!\right]\!\! ,\quad
    \mib R = \left[\!\! \begin{array}{cc} 10^{-2} & 0 \\ 0 & 10^{-2} \end{array} \!\!\right]\!\!.
\end{eqnarray}
Therefore, we have four free parameters $Q_{1,1}$, $Q_{1,2}$, $R_{1,1}$, and $R_{1,2}$ in this system. Note that $\mib Q$ and $\mib T$ are always symmetric matrices from the definitions of (\ref{eq:18}) and (\ref{eq:20}). Other parameters are set as $T_{1,2}=1$, $\eta=10^{-2}$, and $\sigma^2=5\times10^{-2}$. Various values for $T_{1,2}$ are employed to examine the influence of the correlation of the teacher hidden units. We sometimes use $\kappa\equiv\arccos{\frac{T_{1,2}}{T_{1,1}}}$, the angle of the teacher weight vectors, instead of $T_{1,2}$.

     In this case, $\theta_{ij}$ and $\mib\Theta_{ij}$ in the inverse of the Fisher information matrix (\ref{eq:14}) can be simplified as
\begin{eqnarray}
    \label{eq:36}
    \theta_{1,1}     &=& \theta_{2,2} = c\sqrt{a},\quad \theta_{1,2} = \theta_{2,1} = -c\sqrt{b}, \nonumber\\
    \mib\Theta_{1,1} &=& d\sqrt{a} \left[\!\! \begin{array}{cc} 2a\{a\-b\}\-bQ_{1,1} & bQ_{1,2} \\ bQ_{1,2} & -bQ_{1,1} \end{array} \!\!\right]\!\!, \nonumber\\
    \mib\Theta_{2,2} &=& d\sqrt{a} \left[\!\! \begin{array}{cc} -bQ_{1,1} & bQ_{1,2} \\ bQ_{1,2} & 2a\{a\-b\}\-bQ_{1,1} \end{array} \!\!\right]\!\!, \\
    \mib\Theta_{1,2} &=& \mib\Theta_{2,1} \nonumber\\
                  &=& -d\sqrt{b} \left[\!\! \begin{array}{cc} a\{Q_{1,1}\+1\}\-b^2 & aQ_{1,2} \\ aQ_{1,2} & a\{Q_{1,1}\+1\}\-b^2 \end{array} \!\!\right]\!\!,  \nonumber
\end{eqnarray}
where
\begin{eqnarray}
    \label{eq:37}
    & a \equiv \{Q_{1,1}\+1\}^2 \- Q_{1,2}^2,\quad b \equiv 2Q_{1,1}\+1, & \nonumber\\
    & c \equiv \frac{\pi}{2}\ \frac{\sqrt{ab}}{a\-b},\quad d \equiv \frac{c}{a^2\-b^2}.
\end{eqnarray}

     Here we summarize the order of each variable to $N$. Since the length of the input vector $\mib\xi$ is $O(\sqrt{N})$, $x_i$ and $y_i$ are $O(1)$. This guarantees that the arguments of the activation function $g$ are $O(1)$. Therefore, the lengths of the weight vectors, $\sqrt{Q_{ii}}$ and $\sqrt{T_{ii}}$, are $O(1)$. If the direction of the initial $\mib J_i$ is chosen randomly, the size of $R_{ii}$, the correlation between $\mib J_i$ and $\mib B_i$, is $O(1/\sqrt{N})$. The initial numerical values in (\ref{eq:35}) are defined according to these sizes.

     Figure \ref{fig:2} shows the time evolution of the generalization error. In the GD (Fig. \ref{fig:2}a), the plateau was greatly prolonged as the correlation of the teacher weight vectors rose. In NGD (Fig. \ref{fig:2}b), almost no plateau occurred at any $T_{1,2}$ if $\eta$ was set small enough relative to the initial $R_{1,1}$, and the generalization error was exponentially decreased. The plateau periods of Fig. \ref{fig:1}a were measured and are shown in Fig. \ref{fig:2}c, where we defined a plateau as occurring if $\frac{\partial\ln{\epsilon_g}}{\partial\alpha}>-0.0005$. The order of the plateau lengths was about $O(\kappa^{-3})$ in GD.

     Figure \ref{fig:3} shows the trajectories of the order parameters $(R_{1,1},R_{1,2})$ and $(R_{2,1},R_{2,2})$. Because of the symmetry, the latter plots are mirror images of the former. As $R_{1,1}$ is the correlation between the first student and the corresponding teacher, the initial value is almost $0$ and the goal is $1$; $R_{1,2}$ is the correlation between the first student and the not corresponding teacher, and the initial value is almost $0$ and the goal is $T_{1,2}$. Therefore, the target location of the plots are $(1,T_{1,2})$ and $(T_{1,2},1)$, respectively (shown as $\diamondsuit$). The other order parameters $Q_{1,1}$ and $Q_{1,2}$ are not shown. In the case of GD (Fig. \ref{fig:3}a), the plots start at $\square$, turn back at $\triangle$, then approach $\odot$ (the saddle, as explained in the next section), and finally reach $\diamondsuit$. Actually, the parameters never pass through the same place again because $Q_{1,1}$ and $Q_{1,2}$ are updated. In the case of NGD (Fig. \ref{fig:3}b), the plots start at $\square$ and reach $\diamondsuit$ while avoiding $\odot$.

     We performed a numerical simulation to confirm the dynamics at the above thermodynamics limit. The input dimension was $N=1000$, the teacher weight vectors were set as
\begin{eqnarray}
    \label{eq:38}
    \mib B_1 = \left[\!\! \begin{array}{c} 1 \\ 0 \\ 0 \\ \vdots \\ 0 \end{array} \!\!\right]\!\!,\quad 
    \mib B_2 = \left[\!\! \begin{array}{c} \cos{\kappa} \\ \sin{\kappa} \\ 0 \\ \vdots \\ 0 \end{array} \!\!\right]\!\!,\quad 
\end{eqnarray}
and every initial $\mib J_i$ was randomly and independently chosen from ${\mathcal N}(\mib 0, \mib I/N)$ for each try. Thus, the order parameters $\mib Q$ and $\mib R$ were no longer limited by the restriction of (\ref{eq:34}). The learning was performed using these real weight vectors and the original equations: (\ref{eq:9}) for GD and (\ref{eq:11}) for NGD. Figures \ref{fig:4} and \ref{fig:5} show the time evolution of the generalization error and the trajectories of the order parameters in the same manner as Figs. \ref{fig:2} and \ref{fig:3}, respectively. Both figures support the statistical dynamics well, which suggests the constraint of (\ref{eq:34}) is a rather minor problem and the system retains most of its generality even with that restriction.

\section{Saddle}
     Here, we discuss why NGD is so effective even with a strong correlation between teacher hidden units. We consider the dynamics around the saddle of the generalization error under the conditions of (\ref{eq:33}) and (\ref{eq:34}). This point, where all the differentials of the order parameters are zero and the Hessian matrix is not positive definite nor negative definite, is shown as $\odot$ in Figs. \ref{fig:3} and \ref{fig:5}:
\begin{eqnarray}
    \label{eq:39}
    Q_{1,1} = Q_{1,2} &=& \frac{T_{1,1}+T_{1,2}}{T_{1,1}-T_{1,2}+2}, \nonumber\\
    R_{1,1} = R_{1,2} &=& \frac{T_{1,1}+T_{1,2}}{\sqrt{2\{T_{1,1}-T_{1,2}+2\}}}.
\end{eqnarray}
This saddle is a special point because 1) it corresponds to the goal both in the case of $T_{1,1} = T_{1,2}$ (the teacher is a smaller network: $f\mib{_B}(\mib\xi) = 2g(\mib{B}_1^T \mib\xi)$) and in the case that the student is a smaller network: $f\mib{_J}(\mib\xi) = 2g(\mib{J}_1^T \mib\xi)$, 2) in GD, the plateau occurs around it, and in NGD the student vectors avoid it, 3) it coincides with one of the singular points of the Fisher information matrix since $Q_{1,1}=Q_{1,2}$.
\begin{figure}[t]  
\begin{center} 
    \includegraphics[width=45mm]{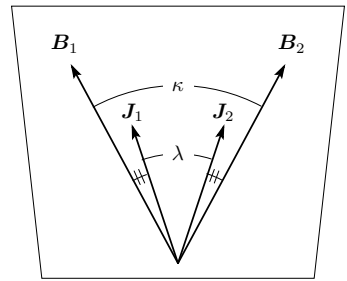}
    \caption{The student weight vectors $\mib J_1$ and $\mib J_2$ belong to the plane made by the teacher weight vectors $\mib B_1$ and $\mib B_2$.}
    \label{fig:6}
\end{center} 
\end{figure}
     We simplify the situation as shown in Fig. \ref{fig:6}; the two student weight vectors belong to the plane made by the two teacher weight vectors. This simplification is useful because we are now interested in how fast the student vectors leave this point for the goals. The correlations are re-parameterized by $\kappa$ and $\lambda$ as
\begin{eqnarray}
    \label{eq:40}
    & T_{1,2} = T_{1,1} \cos{\kappa},\quad Q_{1,2} = Q_{1,1} \cos{\lambda}, & \nonumber\\
    & R_{1,1} = \sqrt{Q_{1,1} T_{1,1}} \cos{\frac{\kappa-\lambda}{2}}, & \nonumber\\
    & R_{1,2} = \sqrt{Q_{1,1} T_{1,1}} \cos{\frac{\kappa+\lambda}{2}}. &
\end{eqnarray}
Now, we have only two free parameters $Q_{1,1}$ and $\lambda$. Since the first derivative of $\lambda$ can be written with $Q_{1,1}$ and $Q_{1,2}$ as
\begin{eqnarray}
    \label{eq:41}
    \frac{\partial\lambda}{\partial\alpha} = \frac{\partial}{\partial\alpha} \arccos{\frac{Q_{1,2}}{Q_{1,1}}}
                          = \frac{Q_{1,2} \frac{\partial Q_{1,1}}{\partial\alpha} - Q_{1,1} \frac{\partial Q_{1,2}}{\partial\alpha}} {Q_{1,1} \sqrt{Q_{1,1}^2 - Q_{1,2}^2}},
\end{eqnarray}
we can formulate the angular velocity of $\lambda$ at $0<\lambda\ll 1$. The term $\eta^2$ included in $\frac{\partial Q_{ij}}{\partial\alpha}$ can be ignored if the learning rate $\eta$ is set small enough.

     The angular velocity for GD is
\begin{eqnarray}
    \label{eq:42}
    \frac{\partial\lambda}{\partial\alpha} = c_1 \lambda \sin^2{\kappa},
\end{eqnarray}
where $c_1 \equiv \frac{4\eta}{\pi\sigma^2} T_{1,1} \{T_{1,1}\{1\-\cos{\kappa}\}\+2\}^{-\frac{1}{2}} \{T_{1,1}\{3\+\cos{\kappa}\}\+2\}^{-\frac{3}{2}}$. We notice that the order of $c_1$ is not greatly changed by $\kappa$. The velocity converges to zero in the first order of $\lambda$. Moreover, it decreases as $\kappa$ decreases. Therefore, this equation supports the simulation results showing that the plateau is prolonged as the teacher correlation rises. The angular velocity for NGD is
\begin{eqnarray}
    \label{eq:43}
    \frac{\partial\lambda}{\partial\alpha} = c_2 \frac{1}{\lambda} \tan^2{\frac{\kappa}{2}},
\end{eqnarray}
where $c_2 \equiv 2\eta$. This velocity diverges to infinity as $\lambda$ goes to zero. Although it decreases as $\kappa$ decreases, this effect would be canceled by $\lambda^{-1}$ near the saddle. Therefore, this equation means that the student weight vectors are repelled by the saddle. In addition, this also supports the simulation results showing that the student weight vectors avoid the saddle and that the plateau does not occur even in the case of strongly correlated teacher hidden units.

\section{Conclusion}
     We have studied the on-line learning of soft committee machines under correlated teacher hidden units. The plateau in GD is largely prolonged at about $O(\kappa^{-3})$ as the correlation of the teacher weight vectors rises, but almost no plateau occurs in NGD with a low learning rate $\eta$ and this does not depend on the correlation. Our analytical results for around the saddle reveal that the NGD avoided the saddle, even though the strong correlation of the teacher weight vectors forced the student weight vectors close to the saddle where the Fisher information matrix is singular.

\end{document}